\documentclass[12pt,preprint]{emulateapj}
\usepackage{amsmath,amssymb,threeparttable}
\defcitealias{Mustill2012}{M12}
\defcitealias{Wisdom1980}{W80}
\defcitealias{Quillen2006}{Q06}

\shorttitle{Gap Clearing in a Collisional Disk}
\begin{document}

\title{Gap Clearing by Planets in a Collisional Debris Disk}

\author{Erika R. Nesvold}
\affil{Department of Physics, University of Maryland Baltimore County 
\\ 1000 Hilltop Circle
\\ Baltimore, MD 21250}
\email{Erika.Nesvold@umbc.edu}

\author{Marc J. Kuchner}
\affil{NASA Goddard Space Flight Center 
\\ Exoplanets and Stellar Astrophysics Laboratory, Code 667
\\ Greenbelt, MD 21230}
\email{Marc.Kuchner@nasa.gov}

\begin{abstract}

We apply our 3D debris disk model, SMACK, to simulate a planet on a circular orbit near a ring of planetesimals that are experiencing destructive collisions. Previous simulations of a planet opening a gap in a collisionless debris disk have found that the width of the gap scales as the planet mass to the 2/7th power ($\alpha=2/7$). We find that gap sizes in a collisional disk still obey a power law scaling with planet mass, but that the index $\alpha$ of the power law depends on the age of the system $t$ relative to the collisional timescale $t_{coll}$ of the disk by $\alpha = 0.32 (t/t_{coll})^{-0.04}$, with inferred planet masses up to five times smaller than those predicted by the classical gap law. The increased gap sizes likely stem from the interaction between collisions and the mean motion resonances near the chaotic zone. We investigate the effects of the initial eccentricity distribution of the disk particles and find a negligible effect on the gap size at Jovian planet masses, since collisions tend to erase memory of the initial particle eccentricity distributions. Finally, we find that the presence of Trojan analogs is a potentially powerful diagnostic of planets in the mass range $\sim1-10 M_{Jup}$. We apply our model to place new upper limits on planets around Fomalhaut, HR 4796 A, HD 202628, HD 181327, and $\beta$ Pictoris.

\end{abstract}

\section{Introduction}
\label{sec:introduction}

Since the first observations of debris disks, spectral energy distributions (SEDs) and resolved images of these disks have often revealed cleared inner regions or gaps. For example, early photometry of the debris disk around Vega showed no excess at wavelengths shorter than $\sim$ 70 microns, indicating a hole in the disk extending out to $\sim$ 80 AU (\citealt{Aumann1984}, but see also \citealt{Su2013}). In the same year, \citet{Smith1984} noted that their coronagraphic images of the $\beta$ Pictoris disk were consistent with a depletion of disk material within 30 AU of the star. We now know that a large fraction of debris disks have inner holes, as indicated by mid- and far-IR photometry \citep{Carpenter2009, Moor2011, Chen2014}.
While some of this radial structure may be inherited from the transitional disk phase when gas plays an important role in the dynamics \citep{Merin2010}, gravitational perturbation by planetary companions could form many of these observed gaps. Collisionless N-body simulations show that planets can create gaps via planetesimal scattering in overlapping resonances. For example, \citet{Roques1994} and \citet{LecavelierdesEtangs1996} used N-body simulations of the $\beta$ Pictoris system to demonstrate that gravitational perturbations by a planet at 20 AU could clear the gap in the disk inferred from the SED. 

More recently, collisionless dynamical models of gap opening by planets have been applied to debris disks that contain planet candidates detected via direct imaging, and used to constrain the properties of the planet. \citet{Quillen2006} used a collisionless model of gap opening to predict the properties of a planet clearing a gap in the debris ring around Fomalhaut. Then, after \citet{Kalas2008} detected a candidate planet around Fomalhaut \citep[see also][]{Janson2012, Currie2012}, \citet{Chiang2009} used the observed gap size to constrain its mass. Recently, \citet{Rodigas2014} used collisionless N-body simulations to derive a relationship between ring width and planet mass, yielding upper limits on the masses of planets in several debris disks, in the context of their collisionless model.

These collisionless dynamical models have revealed some of the basic physics of gap clearing. However, collisions between planetesimals can also affect the radial structure of a debris disk. As \citet{LecavelierdesEtangs1996} and \citet{Quillen2006} foreshadowed, planetesimal collisions can affect the structure of a gap in a debris disk carved by a planet. \citet{Wyatt2005a} showed how collisions between dust grains can effectively open a gap in a dust cloud when the collision time is shorter than the Poynting-Robertson time. Planetesimal collisions have also been shown to create time-evolving radial structures in debris disks by producing dust in response to the formation of large bodies \citep{Kenyon2004, Kennedy2010}. Recent numerical models that incorporate both collisions and dynamics \citep{Stark2009} demonstrated that collision rates among grains in a dust disk are enhanced in mean motion resonances with a perturbing planet.

Since collisions can affect the radial structure of a disk, we need to model the planetesimal collisions and dynamics simultaneously to build a more accurate picture of gaps and inner holes in debris disks. Therefore, we investigated the effects of fragmenting collisions on the evolution of a planetesimal disk containing a planet using our 3D collisional algorithm SMACK \citep{Nesvold2013}. Previous studies of gap opening by planets in debris disks have not modeled both planetesimal collisions and dynamics in 3D.

In particular we re-examine the gap width-planet mass relationship derived from collisionless models \citep[e.g. by][]{Quillen2006, Chiang2009, Rodigas2014} and based on resonance overlap. A planet orbiting in a debris disk is surrounded by a ``chaotic zone'' of unstable orbits caused by overlapping mean motion resonances \citep{Chirikov1979, Wisdom1980}. Planetesimals entering the chaotic zone are scattered onto highly eccentric orbits after $\sim$1000 orbital periods, creating an underdensity of material around the planet's orbit. \citet[][hereafter \citetalias{Wisdom1980}]{Wisdom1980} analyzed the restricted three-body problem to derive a relationship between the size of the chaotic zone and the planet's mass for low-eccentricity particles with semi-major axes close to that of a planet on a circular orbit:
\begin{equation} \label{eq:wisdom} da/a = k \mu^{2/7}, \end{equation}
where $da/a = (a_g - a_p)/a_p$, $a_g$ is the semi-major axis of the outer edge of the chaotic zone,  $a_p$ is the planet's semi-major axis, $k$ is a constant, and $\mu$ is the ratio of the planet's mass to the stellar mass. \citetalias{Wisdom1980} derived a coefficient of $k =1.3$. \cite{Duncan1989} and \cite{Murray1999} derived similar 2/7 scaling laws with coefficients of $k =1.49$ and 1.57, respectively, using an iterated encounter map. \citet{Gladman1993}, however, examined the stability of the full three-body problem of two planets and a star, and found that the stability of the system depended on $\mu^{1/3}$, using analytic derivation and collisionless numerical simulations. \citet{Chiang2009} found a 2/7 law with a coefficient of $k = 2.0$ for a disk of parent bodies and small dust grains using collisionless N-body integrations that included the dynamical perturbations of stellar radiation. 

In this paper we use SMACK to investigate the effects of collisions on the form and parameters of this power law as applied to the distribution of planetesimals $\gtrsim 1$ mm, i.e., planetesimals observed with ALMA and other sub-mm telescopes. Section (\ref{sec:smack}) of this paper describes the simulations we performed. In Section (\ref{sec:results}) we present our results and analysis. In Section (\ref{sec:procedure}) we discuss the implications for observers, and in Section (\ref{sec:summary}), we summarize our results.

\section{Collisional Simulations}
\label{sec:smack}

The Superparticle-Method Algorithm for Collision in Kuiper Belts (SMACK) uses the N-body integrator REBOUND \citep{Rein2012} and a superparticle approximation to simulate the dynamical and size distribution evolution of a disk of planetesimals in 3D as they experience fragmenting collisions \citep{Nesvold2013}. Each body in the N-body integrator represents a superparticle, a cloud of planetesimals with the same location and trajectory but a range of masses, characterized by a size distribution. When a collision between superparticles is detected, SMACK replaces the parent superparticles with daughter superparticles whose velocities and size distributions statistically represent the outcome of the planetesimal collisions during the interval since the last superparticle encounter. Any number of planets can also be included in the N-body integrator, which treats the superparticles as massless test particles of finite size. Radiative forces are not included, so the current version of SMACK is best suited to modeling planetesimals $\gtrsim1$ mm, which is appropriate for molding data from ALMA and other (sub)mm telescopes.

Since \citet{Nesvold2013}, we have updated the crushing law SMACK uses to calculate collisional outcomes. We now use the algorithm described by \citet{Leinhardt2012} for collisions in the catastrophic and super-catastrophic disruption regimes to calculate the size of the largest fragment and the fragment size distribution. The catastrophic disruption regime is defined by collisional energies such that the mass of the largest fragment is half the mass of the original planetesimal. At higher collisional energies, in the super-catastrophic regime, the mass of the largest fragment is smaller than half the mass of the original planetesimal, and decreases with increasing collision energy. The fragment size distribution has the form of a power-law with an empirically fit index of $-3.85$. 

To measure the effects of the change in crushing law between this work and \citet{Nesvold2013}, we ran two simulations of a disk with no planet. The initial parameters of the disks are listed in Table (\ref{tab:initial}). The initial optical depth of the disks was $10^{-2}$. We ran each simulation for $10^7$ yr, using the \citet{Nesvold2013} crushing law for one disk and the updated crushing law for the other. We found that at $10^7$ yr, the total brightness of the disk simulated with the new crushing law was 10\% greater than the disk with the old crushing law. However, any morphological variation there might have been between the two simulations was less than the Poisson noise, so we did not observe it.

To measure the effects of collisions on the gap law, we ran several SMACK simulations of a ring of planetesimals orbiting a solar-mass star. To each system we added a planet with zero eccentricity at a semimajor axis of 50 AU. Because each system had a single planet, and the disk had no gravitational influence on the planet, the planet remained fixed in a circular orbit for the entire simulation. Table 1 lists the initial conditions of the planetesimal ring. The size distribution in each superparticle varies during the simulation, and is not generally a power law. However, each superparticle was assigned an initial power law size distribution,
\begin{equation} n(s) dn = C s^{-q} ds, \end{equation}
where $n$ is the number of planetesimals with diameter between $s$ and $s+ds$. We used a planetesimal size range of $1-100$ mm for the superparticles and set the index of the power law to be $q =3.5$ \citep{Dohnanyi1969}. We calculated $C$ such that the initial face-on optical depth of the disk was $\tau_0 = 10^{-4}$, $10^{-3}$, or $10^{-2}$. When calculating the initial optical depth, we extrapolated the size distributions of the superparticles down to 1 $\mu$m. The simulations covered ten different planet masses, ranging from 0.003 $M_{Jup}$ to 100 $M_{Jup}$ in logarithmic steps, for a total of 30 simulations. Each simulation ran for $10^7$ yr, with the longest simulation requiring $\sim20$ hours of wall clock time when parallelized with OpenMP/MPI on 48 cores on the NASA Center for Climate Simulation's (NCCS) Discover cluster. The optical depth of each disk decreased by a factor of $\sim10$ in $10^7$ yr as collisions ground the planetesimals into dust grains, which were removed from the system. For example, the disk around $\beta$ Pictoris (age 12 Myr, vertical optical depth $\tau_{\perp} = 10^{-4}$ at 10 AU) would be most similar to the SMACK simulations with $\tau_{0} = 10^{-3}$.  

\begin{table} [!ht]
	\begin{tabular} {l c}
	Parameter		 			& Value \\
	\tableline
	Semi-Major Axis (AU) 		& 50-130 \\
	Eccentricity 				& 0.0-0.2 \\
	Inclination 				& 0.0-0.1 \\
	Longitude of Ascending Node 	& 0-$2\pi$ \\
	Argument of Periapsis 		& 0-$2\pi$ \\
	Mean Anomaly 				& 0-$2\pi$ \\
	Size Distribution Index		& 3.5 \\
	Planetesimal Size Range		& 1-100 mm \\
	Vertical Optical Depth		& $10^{-4}, 10^{-3}, 10^{-2}$\\
	\end{tabular}
\caption{Initial conditions of the superparticles for the simulations described in Section \ref{sec:smack}. Each orbital parameter is uniformly distributed within the range listed.}
\label{tab:initial}
\end{table}

SMACK is subject to numerical noise arising from the finite size of the superparticles. Numerical heating can cause the eccentricities of the superparticles to increase artificially. Numerical viscosity can cause a narrow ring of superparticles to spread on an artificially short timescale. However, selecting a small enough superparticle size can increase the timescales for numerical heating and numerical viscosity to greater than the simulation time. 

We used the techniques described in \citet{Nesvold2013} to choose a superparticle radius appropriate for a simulation time of $10^7$ yr. We ran several simulations of a planet-less ring with the parameters listed in Table (\ref{tab:initial}) with different superparticle sizes. We then plotted the mean eccentricity of the ring vs. time and compared the curves for different superparticle sizes. The eccentricity damping curves began to converge at a superparticle size of 0.1 AU, indicating that numerical heating is not a significant source of noise within $10^7$ yr for superparticle sizes $\lesssim$ 0.1 AU. We then used Eq. (21) of \citet{Nesvold2013} to calculate the expected widening of the ring due to numerical viscosity. We found that the planet-less ring will spread by $1.7\%$ in $10^7$ yr due to numerical viscosity, an acceptable amount, so we chose a superparticle size of 0.1 AU for all the SMACK simulations described in this paper.

SMACK simulates a system of discrete particles and is therefore also subject to Poisson noise. For each simulation, we used N = 10,000 superparticles and recorded the orbital elements and grain size distributions of each superparticle every $10^4$ yr. We averaged together the outputs from the last 50 timesteps ($5\times10^5$ yr) to mitigate the Poisson noise. As a result, in the radial distribution of the superparticles in each simulation with a bin size of 0.5 AU, the average number of superparticles per bin was 998, with a corresponding Poisson noise level of $\sim 3\%$. We then calculated the face-on surface brightness of the disk at 850 $\mu$m, assuming spherical blackbody grains and a stellar luminosity of $L_{\odot}$. 

\section{Results}
\label{sec:results}

\begin{figure*} [!ht]
	\centering
	\includegraphics[scale=0.46]{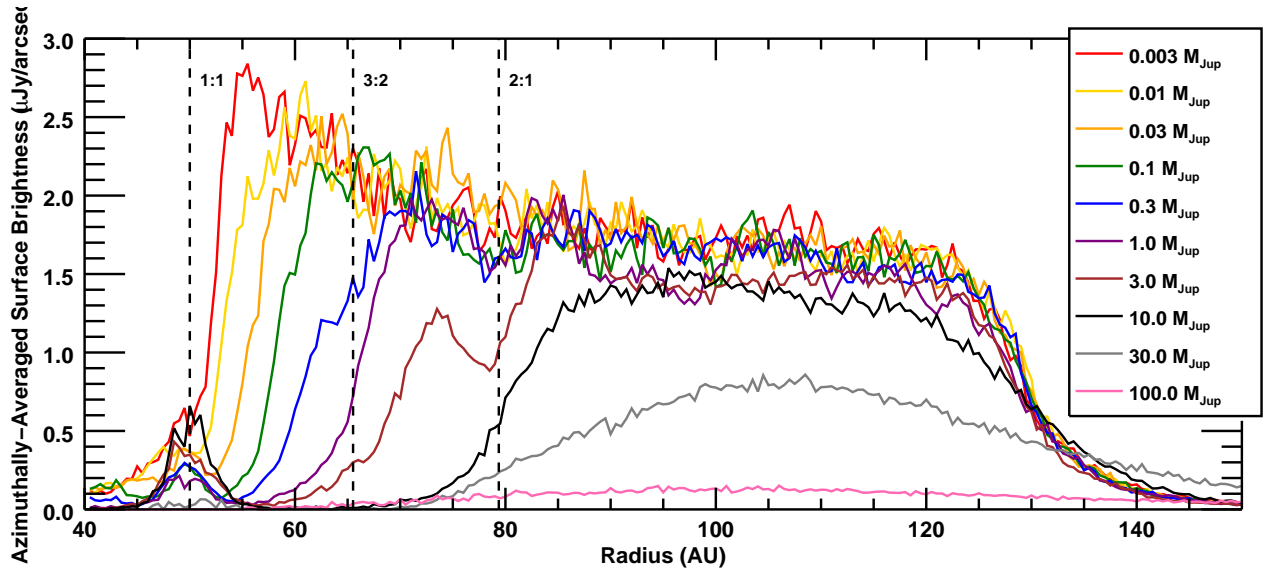}
	\caption{Azimuthally-averaged radial surface brightness at 850 $\mu$m at $10^7$ yr for each of the ten simulations with initial ring optical depth $\tau_0 = 10^{-3}$. The 1:1, 3:2, and 2:1 mean motion resonances are indicated with vertical dashed lines.}
	\label{fig:histall}
\end{figure*}

Fig. (\ref{fig:histall}) shows the azimuthally-averaged radial surface brightness for each simulation with initial optical depth $\tau_0 =10^{-3}$ and a resolution of 1 AU. It illustrates that that larger mass planets clear wider gaps in the disk as expected from Eq. (\ref{eq:wisdom}). Moreover, three new phenomena appear in Fig. (\ref{fig:histall}) that Eq. (\ref{eq:wisdom}) does not predict.

First, each simulation shows a peak in surface brightness at the planet's semimajor axis of 50 AU representing planetesimals trapped in a 1:1 mean motion resonance (MMR) with the planet. We will discuss these Trojan asteroid analogs in more detail in Section (\ref{sec:trojans}).

Secondly, Fig. (\ref{fig:histall}) shows a peak in surface brightness between the 3:2 and 2:1 MMRs for the $M_{p}=3 M_{Jup}$ simulation. This discrete peak between the 3:2 and 2:1 MMRs is not captured in the \citetalias{Wisdom1980} law, which is a continuous approximation to the size of the resonance overlap region. Many models of debris disk images have invoked over densities associated with dust trapped in first-order MMRs with a planet \citep[][etc.]{Kuchner2003, Wyatt2003, Reche2008}. Though migrating planets or migrating dust may lead to such a configuration, the simple static case that we model here produces the opposite effect: planetesimals are depleted in the 3:2 and 2:1 first-order resonances. 

Finally, the sharpness of the outer edge of each disk is very similar for smaller mass planets, but for planets 10 $M_{Jup}$ and higher the outer edge broadens as planet mass increases. This broadening arises because the larger mass planets are stirring the planetesimals to higher eccentricities faster than collisions can damp the planetesimal eccentricities. This behavior was also observed by \citet{Quillen2006}, \citet{Quillen2006a}, and \citet{Chiang2009} and quantified by \citet{Rodigas2014}.

\subsection{Gap Size vs. Planet Mass}
\label{sec:mass}

To use a relationship like Eq. (\ref{eq:wisdom}) to constrain the masses of planetary perturbs, we need to consider how the location of the gap edge is measured. Previous methods for locating the gap edge have considered particle lifetimes \citep{Quillen2006a, Rodigas2014}, eccentricity evolution \citep{Duncan1989, Mustill2012}, or the width of the remaining particle ring \citep{Chiang2009}. We used the half-maximum radius defined by \cite{Chiang2009}, which is easy to compare with observations of resolved disks. The inner edge of the disk, $r_g$, is defined as the smallest radius at which the radial surface brightness profile of the disk disk reaches half its maximum value. We calculated the relative radial size of the gap, $dr/r = (r_g - r_p)/r_p$, for each simulation, combining the last 50 output timesteps as described in Section (\ref{sec:results}). Because the planet's eccentricity is zero in each of our simulations, $dr/r$ is equivalent to $da/a$ from an observational standpoint.

\begin{figure} [!ht]
	\centering
	\includegraphics[scale=0.4]{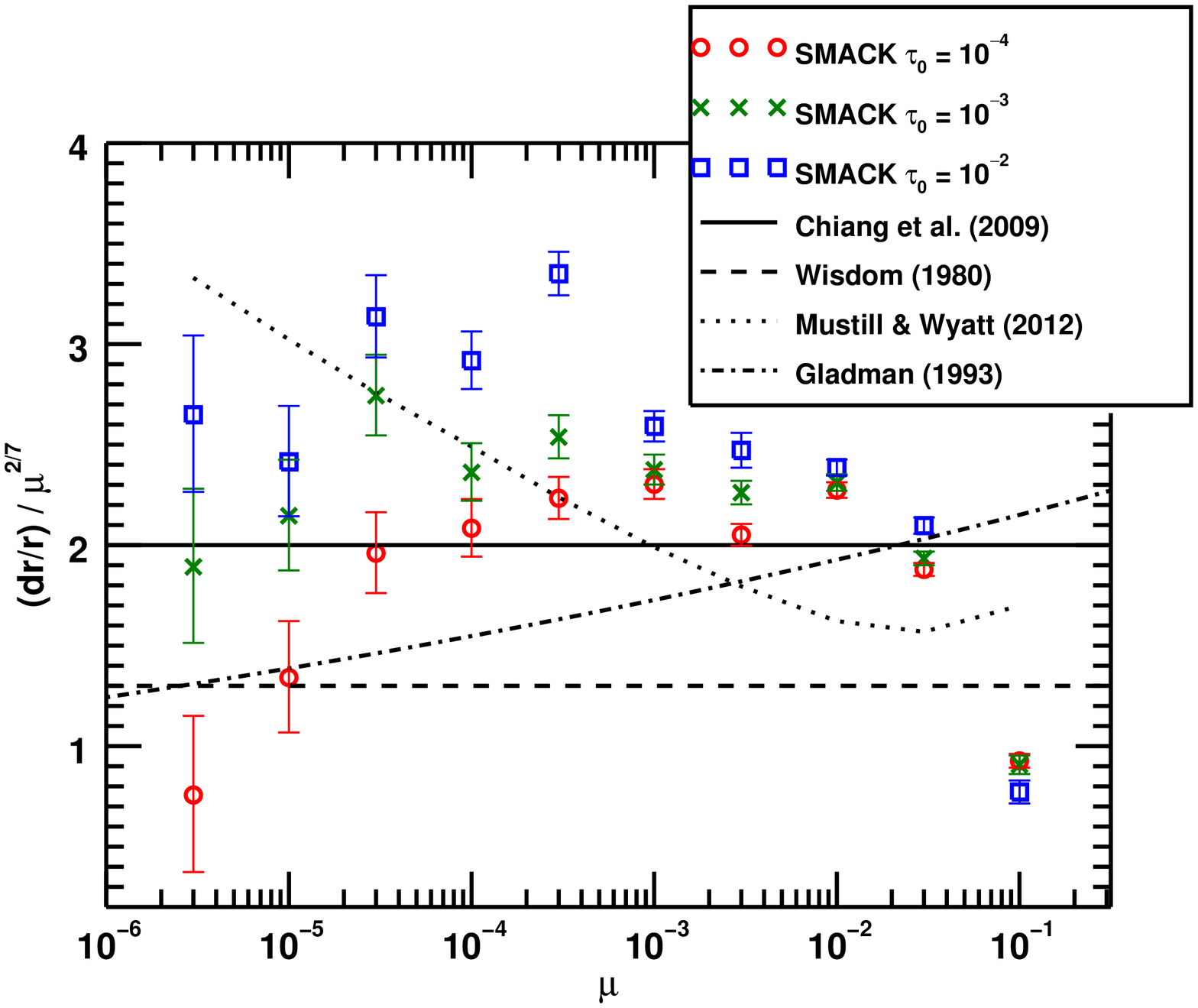}
	\caption{Relative gap size at $10^7$ yr vs. planet-to-star mass ratio for each SMACK simulation, compared with the analytic prediction of \citetalias{Wisdom1980} and the numerical simulations of \citet{Gladman1993}, \citet{Chiang2009} and \citet{Mustill2012}. A $\mu^{2/7}$ dependence has been divided out. The optical depths listed are the initial optical depths $\tau_0$ of each simulation, which decreased by a factor of $\sim10$ during the $10^7$ yr simulation. The error bars include both the negligible Poisson noise and the expected spreading due to numerical viscosity (see Section \ref{sec:smack}), but the uncertainty is dominated by the finite size of the superparticles.} 
	\label{fig:gaplaw}
\end{figure}

Fig. (\ref{fig:gaplaw}) shows our results compared with the 2/7 laws of \citetalias{Wisdom1980} and \citet{Chiang2009} and the 1/3 law of \citet{Gladman1993}, as well as the eccentricity-dependent law of \citet{Mustill2012}, discussed further in Section (\ref{sec:eccentricity}). In Fig. (\ref{fig:gaplaw}) we have divided out the $\mu^{2/7}$ dependence. This figure summarizes three sets of simulations with various initial optical depths, $\tau_0 = 10^{-4}, 10^{-3},$ and $10^{-2},$ measured perpendicular to the disk plane at 100 AU. The results from each SMACK simulation are indicated by symbols, grouped according to their initial optical depths. The solid line and dashed line indicate the predictions of \citet{Chiang2009} and \citetalias{Wisdom1980}, respectively, while the dotted line indicates the particle eccentricity-dependent prediction of \cite{Mustill2012}, which we discuss in more detail in Section (\ref{sec:eccentricity}).

Given the numerical viscosity, our results agree reasonably well with the collisionless model of \cite{Chiang2009} at initial optical depths of $\tau_0 = 10^{-4}$ (indicated by the circles in Fig. \ref{fig:gaplaw}), where collisions are rare, and for planet-to-star mass ratios of $\mu \gtrsim 10^{-4.5}$ (lower-mass planets at this optical depth do not finish opening a gap within $10^{7}$ yr). However, the gaps created in our simulations are wider by up to $70\%$ for disks with higher $\tau_0$ (squares and x's in Fig. \ref{fig:gaplaw}) and correspondingly higher collision rates, a phenomenon that has not previously been reported. 

The relative gap sizes in Fig. (\ref{fig:gaplaw}) begin to decrease relative to the $\mu^{2/7}$ for $\mu \gtrsim 10^{-2}$. In fact, we did find that $dr/r$ saturates at a maximum value of $dr \approx 0.8 r$. Companions with mass ratios $\mu \gtrsim 10^{-2}$ do not continue to produce larger gaps, but they do cause the ring to spread, as shown in Fig. (\ref{fig:histall}). At these mass ratios for a solar-mass star, the companion would have mass $\gtrsim 10$ $M_{Jup}$, almost in the range of brown dwarfs. 

\subsection{Time Dependence}
\label{sec:time}

\citet[][hereafter \citetalias{Quillen2006}]{Quillen2006} argued that inelastic collisions between particles in a disk cause the particle distribution to diffusive from an initially sharp ring edge. Therefore, to maintain a low particle density within a gap, the dynamical removal timescale of particles by the overlapping resonances near the planet must be shorter than the collision timescale in the ring. \citetalias{Quillen2006} used this argument to propose a minimum planet mass, below which a planet would not be able to open a gap in the Fomalhaut disk. However, \citetalias{Quillen2006} did not consider the destructive effects of collisions, which can remove mass from the system by collisionally grinding dust grains down to the blowout size. Our SMACK simulations include this collisional grinding.

The collision timescale for a planetesimal in a disk can be approximated by 
\begin{equation} \label{eq:tcoll} t_{coll} = t_{per}/4 \pi \tau_{eff}, \end{equation}
where $t_{per}$ is the orbital period and $\tau_{eff}$ is the effective optical depth of the belt \citep{Wyatt2009}. We estimated the removal timescale in our simulations as a function of $\mu$ as
\begin{equation} t_{rem} = t_{2/3} T_p = 0.23 \mu^{-0.84} T_p , \end{equation}
$T_p$ is the period of the planet and $t_{2/3}$ is the lifetime in planet orbits of particles with an initial semimajor axis two-thirds of the way between the planet's semimajor axis and the chaotic zone boundary, a relationship derived numerically by \citet{Quillen2006a}. \citetalias{Quillen2006}'s nondestructive collision argument would predict that in systems with $t_{coll} \lesssim t_{rem},$ the planet would be unable to open a gap in the disk. 

For the $\tau_0 = 10^{-2}$ SMACK simulations, $t_{coll} < t_{rem}$ for all systems with $\mu \lesssim 10^{-2}$. For $\tau_0 = 10^{-3}$, $t_{coll} < t_{rem}$ for all systems with $\mu \lesssim 10^{-3}$. However, every one of the planets in our 30 SMACK simulations was able to open a gap. In SMACK simulations including destructive planetesimal collisions, there is no minimum planet mass criterion for opening a gap in a disk. Indeed, destructive collisions will eventually create a gap even without the presence of a planet, starting in the region with the shortest collision time.

\begin{figure} [!ht]
	\centering
	\includegraphics[scale=0.28]{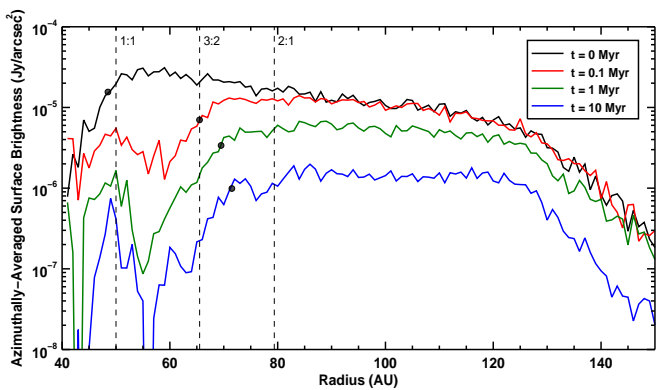}
	\caption{Azimuthally-averaged surface brightness at 850 $\mu$m of the SMACK simulation with mass ratio $\mu = 10^{-2.5}$ and initial optical depth $\tau = 10^{-3}$ at four different times. Each curve except $t=0$ is an average over 10 output timesteps ($10^5$ yr). The dashed vertical lines indicate the locations of the 1:1, 2:1, and 3:2 MMRs. The black circles indicate the half-maximum radius of each distribution.}
	\label{fig:histtime}
\end{figure}

We investigated the time evolution of the gaps in our simulations. Fig. (\ref{fig:histtime}) shows the time evolution of the radial surface brightness profile of the SMACK simulation with $\mu = 10^{-2.5}$ and $\tau_0 = 10^{-3}$. As our simulations do not include a reservoir of large particles to replenish the disk, the total surface brightness of the disk decreased over time. Beyond $\sim70$ AU, the surface brightness distribution evolved more or less homologously. But interior to $\sim70$ AU, the surface brightness continued to change shape throughout the simulation. The gap is cleared of most material by 1 Myr, which would be approximately 100 particle lifetimes according to the power law of \citet{Quillen2006a}. After 1 Myr, planetesimals in the 2:1 resonance continue to collisionally erode, creating a deficit in the surface brightness profile around 79 AU. 

\begin{figure} [ht!]
	\centering
	\includegraphics[scale=0.4]{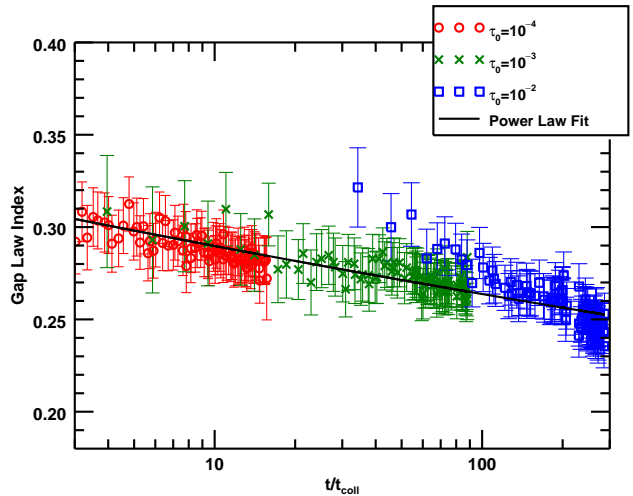}
	\caption{Index $\alpha$ of the gap size power law fit to the SMACK simulations vs. simulation time measured in units of initial collision time. The black line indicates our power law fit (Equation \ref{eq:index}).}
	\label{fig:index}
\end{figure}

To quantify the long-term evolution of the gap shape, we measured the time evolution of the gap size vs. $\mu$ relationship by measuring the relative gap size for each simulation at intervals of $10^5$ yr. We fit a power law of the form $dr/r = k \mu^\alpha$ to these gap sizes at each time step. To observe how this power law evolved, we plotted the index $\alpha$ against the age of the system measured in units of the collision timescale (Fig. \ref{fig:index}). We estimated the collision timescale, $t_{coll}(t)$, as a function of time using Equation (\ref{eq:tcoll}), using is the measured vertical optical depth $\tau(t)$ at time $t$. In this way, we were able to compare the time evolution of the power law across different SMACK simulations directly, despite varying initial optical depths. 

The index decreased slightly over time. We fit a power law to our results and found
\begin{equation} \label{eq:index} \alpha (t/t_{coll}) = (0.318  \pm 0.002) (t/t_{coll})^{-0.041 \pm 0.001}. \end{equation}
We then fixed $\alpha$ to the power law in Equation (\ref{eq:index}) and fit a power law of the form $dr/r = k \mu^{\alpha(t/t_{coll_0})}$ to the gap widths to determine the coefficient $k$. We found that $k$ shows no discernible trend over $t/t_{coll_0}$, remaining within the range $k = 2.06 \pm 0.17$. In general, collisions increase the width of the gap over time faster for smaller-mass planets, creating a shallower gap law.

\subsection{Dependence on Initial Conditions}
\label{sec:eccentricity}

Since collisions tend to erase the memory of the exact initial state of the system, we expect that our simulations will be relatively insensitive to the initial conditions. However, the widths of mean motion resonances can vary with particle eccentricities and inclinations, as do collision velocities. We need to explore how our results depend on the initial eccentricity and inclination distributions of the planetesimals. 

\begin{figure} [ht!]
	\centering
	\includegraphics[scale=0.4]{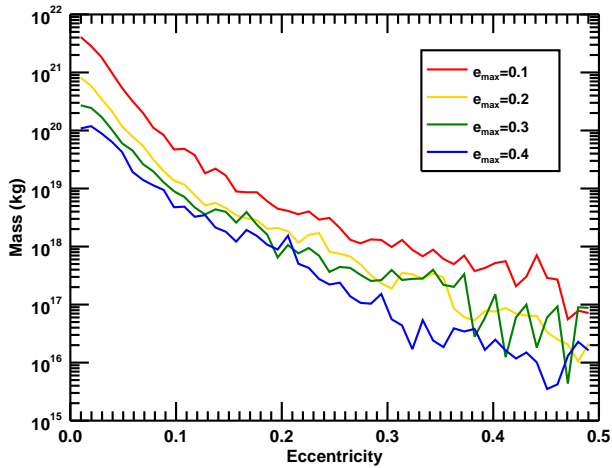}
	\caption{Planetesimal eccentricity distributions at $10^7$ yr for four SMACK simulations with $\mu=10^{-3}$, $\tau_{0}=10^{-2}$, and varying initial maximum eccentricities. Higher initial eccentricities result in accelerated mass loss, but the final eccentricity distributions at $10^7$ yr are remarkably independent of the initial conditions.}
	\label{fig:eccdist}
\end{figure}

To explore how the initial eccentricity distribution affect our simulations, we ran four simulations of a disk and planetary system with $\mu=10^{-3}$ and $\tau_{0}=10^{-2}$. The initial conditions of the simulations were the same as those listed in Table (\ref{tab:initial}), except that the eccentricities were uniformly distributed between 0 and some maximum eccentricity $e_{max}=0.1,0.2,0.3$ or 0.4, and the inclinations were uniformly distributed between 0 and $e_{max}/2$ \citep{Krivov2005}. Collisions damped the eccentricities of the planetesimals during the simulations, producing very similar eccentricity distributions at $10^7$ yr for each simulation, as shown in Fig. (\ref{fig:eccdist}). (But recall that for mass ratios $\mu > 10^{-2}$, the planetesimals are scattered to high eccentricities faster than collisions can damp the planetesimal eccentricities, broadening the outer edge of the ring, as seen in Fig. \ref{fig:histall}). 

\begin{figure} [ht!]
	\centering
	\includegraphics[scale=0.4]{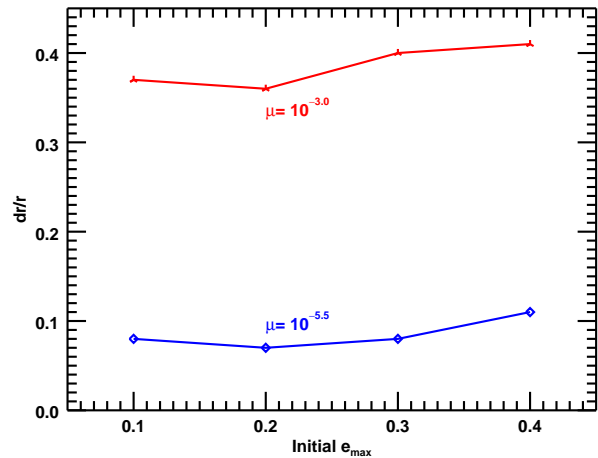}
	\caption{Measured $dr/r$ at $10^7$ yr for simulations with an Earth- or Jupiter-mass planet in a disk with initial optical depth $\tau_0=10^{-2}$ vs. maximum initial eccentricity $e_{max}$. Despite the very similar final planetesimal eccentricity distributions, gaps in simulations with $e_{max} \gtrsim 0.3$ are larger by $\sim10\%$ for simulations with a Jupiter-mass planet and $\sim60\%$ for simulations with an Earth-mass planet.}
	\label{fig:eccgaploc}
\end{figure}

Fig. (\ref{fig:eccgaploc}) plots the measured relative gap size $dr/r$ versus $e_{max}$ at $10^7$ yr for each of the four simulations with a Jupiter-mass planet ($\mu = 10^{-3}$) as well as four simulations with an Earth-mass planet ($\mu = 10^{-5.5}$). The gap size remains roughly consistent until $e_{max} \gtrsim 0.3$. For $\mu = 10^{-3}$, the relative gap size is $\sim10\%$ larger in the $e_{max} = 0.4$ simulation than the $e_{max} = 0.2$ simulation, while for $\mu = 10^{-5.5}$, the relative gap size increases by $\sim60\%$ from the $e_{max}=0.2$ to the $e_{max} = 0.4$ simulations. We conclude that in simulations with $e_{max} < 0.3$, the gap is not affected by the initial eccentricity of the planetesimals. More simulations are needed to determine the relationship between non-zero planet eccentricities and gap size, but our time-dependent gap law (Equation \ref{eq:index}) is applicable to a significant subset of debris disks exhibiting gaps.

Other authors have explored the effect of particle eccentricity on gap width, without the benefit of collision models. \cite{Bonsor2011a} demonstrated with N-body simulations that the width of the chaotic zone increases for higher-eccentricity particles. \citet[][hereafter \citetalias{Mustill2012}]{Mustill2012} used N-body simulations and the iterated encounter map of \cite{Duncan1989} to demonstrate that chaotic zone width is independent of particle eccentricity in the low-eccentricity regime, but increases with eccentricity for moderate particle eccentricities. We plot \citetalias{Mustill2012}'s gap width law in Fig. (\ref{fig:gaplaw}):
\begin{equation} \label{eq:mustill} da/a = 1.8 e_{particle}^{1/5} \mu^{1/5}. \end{equation}
Equation (\ref{eq:mustill}) holds for particles with eccentricity greater than a critical eccentricity
\begin{equation} \label{eq:critecc} e_{crit} \approx 0.21 \mu^{3/7}, \end{equation}
but less than a maximum eccentricity
\begin{equation} \label{eq:ejectecc} e_{eject} = 2.1 \mu^{1/4}, \end{equation}
beyond which particles are on planet-crossing orbits and are removed via close encounters rather than chaotic diffusion. In the \citetalias{Mustill2012} picture, particles with $e < e_{crit}$ obey the 2/7 scaling law of \citetalias{Wisdom1980}. 

However, the \citetalias{Mustill2012} does not seem to be supported by our simulations. We measured the mean eccentricity of the superparticles in each of our 30 simulated disks at the end of $10^{7}$ yr. The mean eccentricities range from 0.6-0.12. Every disk had a mean eccentricity higher that the critical eccentricity derived by \citetalias{Mustill2012} and less than the maximum ejection eccentricity (see Equations \ref{eq:critecc} and \ref{eq:ejectecc}). For Fig. (\ref{fig:gaplaw}), we calculated the gap size predicted by \citetalias{Mustill2012} (Equation \ref{eq:mustill}) using the mean eccentricities of the superparticles; except at the highest value for $\mu$, where the eccentricities of the superparticles were stirred to high values by the planet, the weak $e^{1/5}$ eccentricity dependence had negligible effect on the predicted gap size. The gap widths predicted by \citetalias{Mustill2012} yield a poor fit for the SMACK simulations.

The planet's eccentricity can also play an important role in the debris ring, primarily by forcing an eccentricity on the debris disk particles. However, \citet{Quillen2006a} found that the width of a first-order resonance for a particle with zero free eccentricity near an eccentric planet is the same as for a zero-eccentricity particle near a zero-eccentricity planet, and that the dynamics of these two classes of particles are similar. Their collisionless numerical simulations showed that for a planet with eccentricity $< 0.3$, the collisionless gap law is independent of planet eccentricity. More simulations are needed to explore the relationship between non-zero planet eccentricities and gap size in the presence of collisions.

\subsection{Trojan Planetesimals}
\label{sec:trojans}

\begin{figure} [ht!]
	\centering
	\includegraphics[scale=0.4]{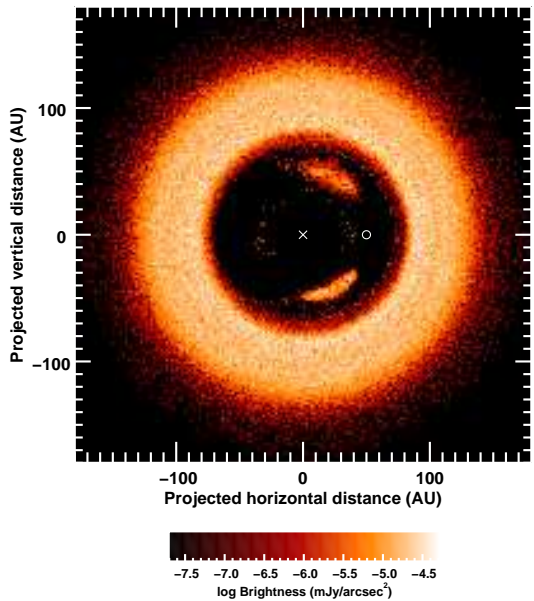}
	\caption{Simulated image at 850 $\mu$m of the SMACK planetesimal ring with $\tau_0 = 10^{-3}$ and planet with $\mu = 10^{-2}$. The brightness of the ring is averaged over the last $5\times10^5$yr in a frame co-rotating with the planet. The star is indicated by a white x. The planet is indicated by a white circle.}
	\label{fig:trojanimage}
\end{figure}

The radial surface brightness profiles in Fig. (\ref{fig:histall}) show a peak in brightness near the semi-major axis of the planet at 50 AU, which varies with planet mass. This peak represents planetesimals trapped in the 1:1 resonance with the planet, analogous to the Trojan asteroids orbiting at Jupiter's L4 and L5 Lagrange points. Fig. (\ref{fig:trojanimage}) shows a simulated image of the $\tau_0 = 10^{-3}, \mu = 10^{-2}$ SMACK simulation at $10^7$yr and a wavelength of 850 $\mu$m. The x indicates the position of the star, and the circle indicates the position of the planet. The image shows two clumps of material in leading and trailing Lagrange points. Though many models of clumps in debris disks have relied on density enhancements associated with MMRs \citep[][etc.]{Kuchner2003, Wyatt2003, Reche2008}, we find in our simulations (which do not include planet or dust migration) that MMRs are depleted in planetesimals except for the 1:1 MMR.

The presence of these Trojan planetesimals could be a useful diagnostic for the mass of the perturbing planet; the Trojan population peaks at $\mu \approx 10^{-2}$. To quantify this phenomenon, we define the relative radial surface brightness of these Trojans, $b_{rel}$, as the ratio of their peak radial surface brightness, $b_{peak}$, to the maximum radial surface brightness of the ring, $b_{ring}$. We measured $b_{peak}$ for each simulation by fitting a Gaussian to the radial surface brightness profile around 50 AU for each simulation. We were unable to accurately fit a Gaussian to the radial surface brightness profiles of the Trojans in the six simulations with $\mu \leq 10^{-5}$ due to confusion with the nearby planetesimals in the ring. In the three simulations with $\mu = 10^{-1.0}$, we did not detect any contributions from planetesimals at 50 AU. 

\begin{figure} [ht!]
	\centering
	\includegraphics[scale=0.4]{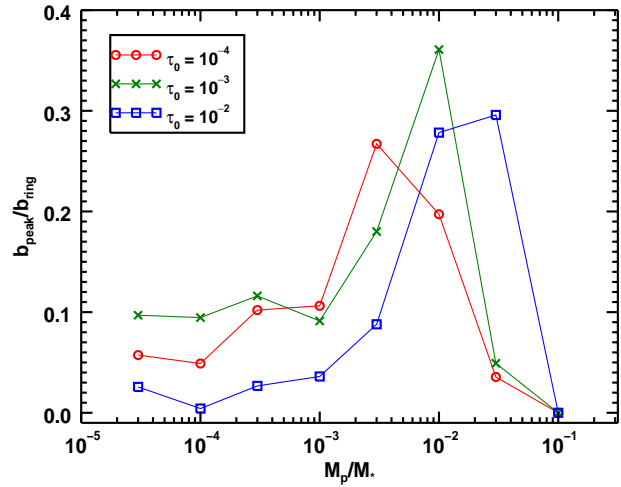}
	\caption{Relative peak radial surface brightness $b_{rel}$ vs. planet-to-star mass ratio $\mu$ for a subset of the SMACK simulations with various initial optical depths $\tau_0$. For each $\tau_0$, $b_{rel}$ increases with $\mu$ before sharply dropping to zero at $\mu \approx 10^{-1.5}$. The sharp peaks suggest that the presence of Trojan planetesimals could be a powerful diagnostic of planet mass.}
	\label{fig:trojanlaw}
\end{figure}

Fig. (\ref{fig:trojanlaw}) shows a plot of $b_{rel}=b_{peak}/b_{ring}$ vs. $\mu$. The relative brightness of the Trojans increases with increasing planet mass until $\mu \approx 10^{-1.5}$, after which the Trojan population drops to zero. The peak in relative Trojan brightness shifts to larger $\mu$ for increasing $\tau_{0}$.

We examined the superparticles orbiting in the Trojan-like clumps at the end of each simulation and discovered that the initial distributions of their semimajor axes and longitudes relative to the planet's were similar to the final distributions. This indicates that they initially orbited in the 1:1 MMR and were not scattered into it, and that the Trojan population is not replenished over time. The initial conditions of our SMACK simulations included a sharp cutoff of the planetesimals at 50 AU. The 1:1 MMR may be even more heavily populated if the initial ring extended interior to 50 AU. In a future paper, we will investigate the accumulation and evolution of Trojans with SMACK simulations optimized for studying the 1:1 MMR. 

The orbits of bodies in a 1:1 resonance with a planet can be categorized as tadpole orbits or horseshoe orbits. In all of our simulations exhibiting significant populations of Trojan planetesimals, including the system shown in Fig. (\ref{fig:trojanimage}), the Trojan planetesimals had tadpole orbits. Horseshoe orbits are unstable for $\mu > 1/1200$ \citep{Cuk2012}, and all of our simulated disks with substantial Trojan populations are in this regime. The libration width of Trojans on tadpole orbits generally increases as $\mu^{1/2}$ \citep{Murray1999}. Increasing planet mass will increase the collision velocities within the Trojan clumps, which could grind down Trojan planetesimals faster, possibly causing the drop-off in the relative brightness of the Trojans at high $\mu$ in Fig. (\ref{fig:trojanlaw}). 

\section{Application to Observed Debris Disks}
\label{sec:procedure}

Based on our simulations, we propose the following procedure for predicting the mass ratio $\mu$ of a planet on on a circular orbit in a disk, given a measured gap edge of $r_g$, planet distance from the star $r_p$, disk optical depth $\tau$, and age $t$:
\begin{enumerate}
\item Estimate the disk's collisional timescale $t_{coll}$ with Equation (\ref{eq:tcoll}).
\item Calculate the gap law index $\alpha$ with Equation (\ref{eq:index}).
\item Calculate the planet-to-star mass ratio with $\mu=(dr/2.06r)^{1/\alpha}$. 
\end{enumerate}
Alternatively, the planet's distance from the star $r_p$ can be predicted from the planet's mass.

We now demonstrate this procedure for five bright debris rings with central clearings.

\subsection{Fomalhaut}

Fomalhaut is a nearby A3V star with an eccentric debris ring that has been resolved in scattered light \citep{Kalas2005} and in sub-mm emission \citep{Boley2012}. A candidate planet was imaged interior to the ring \citep{Kalas2008, Currie2012, Galicher2013}, but more recent measurements of the object's orbit indicate that it cannot be responsible for the eccentricity or sharp inner edge of the ring \citep{Kalas2013}. Nonetheless, an unseen planet could still have carved the ring.

Fomalhaut has a mass of 1.92 $M_{\odot}$ and an age of 440 Myr \citep{Mamajek2012}, and the debris ring has a vertical optical depth of $1.6\times10^{-3}$ at the inner edge of the ring \citep{Marsh2005}, which \citet{Kalas2013} measured to be at $r_{in} = 136 $ AU. Using Equation (\ref{eq:tcoll}), we find that the collisional timescale at $r_{in}$ is $t_{coll} = 5.7\times10^4$ yr. The gap law we infer for the Fomalhaut system indicates that planet radius depends on planet mass by $1/r_p = 0.0029 m_p^{0.22} + 0.0074,$ where $r_p$ is measured in AU and $m_p$ in $M_{Jup}$. 

\citet{Currie2013a} ruled out any planet with a mass $m_p > 3$ $M_{Jup}$ at a projected separation larger than 45 AU. According to our calculated gap law, this places an upper limit on the relative gap size of $dr/r \approx 0.5$, corresponding to a minimum planet orbital radius of $r_p \approx 90$. This indicates that if a planet is creating the sharp inner edge of the Fomalhaut ring, it is orbiting between 90 AU and $r_{in} = 136$ AU.
 
\subsection{HR 4796}

HR 4796 A is an A0V star with a narrow debris ring with a sharp inner and outer edge in scattered light \citep{Schneider2009, Thalmann2011, Lagrange2012a}. Although the ring exhibits a large gap and a small offset from the star, both of which could indicate the presence of a perturbing planet \citep{Wyatt1999a, Schneider2009, Thalmann2011}, no planet has yet been detected. 

HR 4796 A has a mass of 2.18 $M_{\odot}$ \citep{Gerbaldi1999} and an age of 8 Myr \citep{Schneider2009}, and the debris ring has a vertical optical depth of $5\times10^{-3}$ and an inner edge at 77.5 AU \citep{Lagrange2012a}. The collisional timescale at $r_{in}$ is $t_{coll} = 7.4\times10^3$ yr, and the gap law we infer for the HR 4796 A system indicates $1/r_p = 0.004 m_p^{0.24} + 0.013,$ where $r_p$ is measured in AU and $m_p$ in $M_{Jup}$. 

\citet{Lagrange2012a} ruled out the presence of planets with mass $m_p > 3.5$ $M_{Jup}$ beyond 36.5 AU. This places an upper limit on the relative gap size of $dr/r \approx 0.44$, and a lower limit on planet orbital radius of $r_p \approx 54$ AU, indicating that if a planet is responsible for the gap in the HR 4706 A disk, it is orbiting between 54 and 77.5 AU. 

\subsection{HD 202628}

HD 202628 is a G2V star with a broad, eccentric debris ring, inclined by $\sim 64^{\circ}$ from face-on. Like the Fomalhaut and HR 4796 A debris rings, the HD 202628 ring has a sharp inner edge, indicating the presence of a planet orbiting interior to the ring \citep{Krist2012}. We approximate the face-on optical depth of the ring as $\tau \approx L_{dust}/L{star} = 1.4\times10^{-4}$ \citep{Koerner2010}. The star has an age of 2.3 Gyr and the inner edge of the ring has a semimajor axis of $\sim 158$ AU \citep{Krist2012}. The collisional timescale at the inner edge of the ring is $1.13\times10^6$ yr, so the gap law we infer for the HD 202628 system is $1/r_p = 0.003 m_p^{0.23} + 0.006$. If we assume that a single, planetary-mass ($< 15$ $M_{Jup}$) companion is responsible for sculpting the inner edge of the ring, and that the star has a mass equal to the Sun's, then our gap law implies that the planet's orbital radius lies in the range $r_p \approx 86-158$ AU.

\subsection{HD 181327}

HD 181327 is an F5/6 star in the $\beta$ Pic moving group \citep{Schneider2006} with an age of 12 Myr. HD 181327 harbors a near-circular ring of debris with a sharp inner edge at $r_{in}\approx 31$ AU \citep{Stark2014}. Again, we approximate the face-on optical depth of the ring with $\tau \approx L_{IR}/L_{*} = 2.5\times10^{-3}$ \citep{Stark2014}, from which we infer a collision timescale at $r_{in}$ of $4.94\times10^3$ yr. The mass of the star is 1.36 $M_{\odot}$ \citep{Lebreton2012}, so our inferred gap law indicates $1/r_p= 0.013 m_p^{0.23} + 0.032.$ \citet{Wahhaj2013} ruled out any planets with masses $> 6.1$ $M_{Jup}$ beyond $0\farcs36$. This allows us to place a lower limit of $r_p = 19$ AU on the planet orbital radius. If a planet created the gap in the HD 181327 disk, it orbits between 19 and 31 AU.

No azimuthal dust enhancements produced by Trojan asteroids have been identified in this system. If a planet is responsible for shaping the inner edge of the debris ring, planetesimals trapped in the planet's 1:1 MMR may have been lost or destroyed due to the planet's eccentric orbit or some other process. But the near-zero eccentricity of the ring and our results in Fig. (\ref{fig:trojanlaw}) indicate that the lack of detectable Trojans may imply a further constraint on the mass of the planet. According to Fig. (\ref{fig:trojanlaw}), the lack of Trojans in the HD 181327 system places an upper limit of $\sim3.5$ $M_{Jup}$ on the mass of the potential planet. This smaller mass limit changes the lower limit on $r_p$ only slightly, to 20 AU.

\subsection{$\beta$ Pictoris}

$\beta$ Pictoris is an A5V star with a bright, asymmetric, edge-on debris disk. Mid-infrared imaging of this system \citep{Okamoto2004} has detected planetesimal belts at 6.4 and 16 AU, indicating the possible presence of a planet clearing a gap between the belts \citep{Freistetter2007}. A giant planet, $\beta$ Pic b, has been detected orbiting the star at a projected distance of $\sim8$ AU at multiple wavelengths \citep[e.g.][]{Lagrange2009, Lagrange2010, Quanz2010, Currie2011, Boccaletti2013}. The orbit of $\beta$ Pic b is well-characterized due to its relatively short orbital timescale, and recent observations of the planet indicate a best-fit orbit with a semi-major axis of $\sim9$ AU \citep{Chauvin2012, Lagrange2012}. 

$\beta$ Pic has a mass of 1.75 $M_{\odot}$ \citep{Crifo1997} and an age of 21 Myr \citep{Binks2014}. \citet{Ahmic2009} modeled the $\beta$ Pic disk with an optical depth of $\sim2\times10^{-4}$ at 16 AU. This yields a collisional timescale of $1.9\times10^4$ yr and a gap law of $dr/r = 0.35 m_p^{0.24}$. If the inner edge of the gap coincides with the planetesimal belt at 6.4 AU, then a planet orbiting at $r_p = 9$ AU would have a maximum relative gap size of $dr/r = 0.29$, which allows us to place an upper limit of $\sim0.45$ $M_{Jup}$ on the mass of the planet. Radial velocity measurements place much higher upper limits on the mass of $\beta$ Pic b \citep{Lagrange2012}, and evolutionary models estimate a larger planet mass of $7-9$ $M_{Jup}$ \citep{Lagrange2010, Quanz2010, Bonnefoy2011, Currie2013}.

If we assume that $\beta$ Pic b is solely responsible for the gap between the planetesimal belts at 6.4 and 16 AU and require that the planet is located symmetrically between the belts at 11.2 AU, the planet would have a maximum relative gap size of $dr/r = 0.43$ and an upper mass limit of only $\sim2.3$ $M_{Jup}$. This illustrates the strong dependence of planet mass on the relative gap size. 

Fig. (\ref{fig:trojanlaw}) shows that, given the optical depth of the $\beta$ Pic ring, $\beta$ Pic b is near the optimal mass for collecting a population of Trojans with a high surface brightness. \citet{Dent2014} observed asymmetric densities of CO in the $\beta$ Pic disk that could indicate the presence of clumps of planetesimals trapped in 2:1 or 3:2 resonances with a planet producing gas through collisions. These clumps are located $\sim 85$ AU from the star, and are likely not associated with $\beta$ Pic b but rather a second planet orbiting farther out. If $\beta$ Pic b has trapped a population of collisionally active planetesimals in its 1:1 resonances, the gas produced may be detectable with ALMA. Further modeling of the $\beta$ Pictoris system is needed to determine the likely detectability of Trojans near $\beta$ Pic b.

\subsection{Other Systems}

The collisional gap law described in this paper does not break the degeneracy between planet mass and planet semimajor axis that is present in the classical gap law. However, in systems where both edges of a gap can be observed, we can make the assumption that the planet is orbiting symmetrically within the gap, allowing us to place stricter constraints on the planet mass. For example, in a system whose SED is well-fit by a two-temperature grain model, we can interpret it as a two-belt disk with a measurable gap in between and use the collisional gap law to constrain the mass of a possible planet responsible for clearing the gap \citep{Jang-Condell2014}. Disk images that resolve both sides of a gap in a disk, while rarer, can provide even better constraints on the gap width and planet location, and therefore on planet mass. For example, ALMA images of the broad debris disk around the G2V star HD 107146 indicate a dip in disk surface density which may correspond to a fully-depleted gap of width 9 AU \citep{Ricci2014}. By assuming that a planet created the gap by orbiting symmetrically between the gap edges, \citet{Ricci2014} used the collisional gap law to predict a planet mass of $\approx1.9$ $M_{Earth}$. 

\section{Summary}
\label{sec:summary}

We have used our 3D collisional debris disk model SMACK to simulate the opening of a gap in a ring of planetesimals. First, we updated SMACK to use the algorithm described by \citet{Leinhardt2012} for collisions in the catastrophic and super-catastrophic disruption regimes. Then we ran simulations of a planet orbiting in a disk of planetesimals, varying the planet mass and disk optical depth, for $10^7$ yr.

We find that while the size of a gap opened by a planet in a collisional ring still obeys a power law, the index depends on the age of the system relative to the collisional timescale. Our results indicate that in $10^7$ yr, planets can open gaps up to 62\% wider than previously predicted by analytic derivations or collisionless numerical simulations. Correspondingly, the planet mass we infer for a planet in a gap can be up to five times smaller than the mass predicted by the classical gap laws. \citet{Stark2009} found that collisional destruction of grains is enhanced in mean motion resonances; this interaction between collisions and the mean motion resonances near the chaotic zone is probably responsible for the increased gap sizes we see in our simulations. More simulations are needed to determine whether this collisional widening of the observed gap continues past $10^7$ yr.

We applied our results to the Fomalhaut, HR 4796 A, and HD 202628 systems to constrain the radial distances of possible planets sculpting their debris rings. We also analyzed the $\beta$ Pictoris system and placed an upper limit of 0.45 $M_{Jup}$ on the mass of $\beta$ Pic b based on the planetesimal belt at 6.4 AU detected by \citet{Okamoto2004}, a much smaller upper limit than previous models. The time dependence of the collisional gap law implies that the gaps in these observed disks were narrower in the past, and have widened over time due to the presence of collisions.

We also demonstrated that the initial eccentricity distribution of the planetesimals has a negligible effect in systems with a Jovian mass planet. Collisions damp planetesimal eccentricities in a ring, erasing the initial conditions of the system, though initial planetesimal eccentricities above $e_{max}=0.3$ can leave their signature on the radial surface brightness profile of the ring in the form of widened gaps around a planet, especially for low-mass planets.

Finally, we noted that the Trojan-like planetesimals collected into the planet's 1:1 MMR vary in surface brightness with planet-to-star mass ratio $\mu$. The surface brightness of these Trojans increases with planet mass until $\mu \approx 0.3$, beyond which the planet scatters the Trojans by pumping up their eccentricities. The absence of Trojan planetesimals in the Fomalhaut, HD 202628, and HD 181327 systems may place further constraints on the planet mass for each system, and correspondingly the expected semimajor axis range.

\vspace{1em}

We thank Karl Stapelfeldt, Aki Roberge, Hanno Rein, and Margaret Pan for helpful discussions. Erika Nesvold is supported in part by NASA Planetary Geology and Geophysics grant PGG11-0032. Marc Kuchner is supported in part by the NASA Astrobiology Institute through the Goddard Center for Astrobiology. Additional support for this research was provided by NASA through a grant from the Space Telescope Science Institute, which is operated by the Association of Universities for Research in Astronomy, Inc., under NASA contract NAS 5-26555.

\bibliographystyle{apj}
\bibliography{Gaps}

\newpage

\end{document}